\documentclass[prl,%
 reprint,
superscriptaddress,
 amsmath,amssymb,
 aps,
floatfix,
]{revtex4-1}
\usepackage{enumerate}
\usepackage{ulem}
\usepackage{float}
\usepackage{graphicx}
\usepackage{dcolumn}
\usepackage{bm}
\usepackage[dvipsnames]{xcolor}
\usepackage{wrapfig}
\usepackage[left]{lineno}

\begin{document}

\preprint{APS/123-QED}

\title{
\textcolor{blue}{
Destabilisation of local magnetic anisotropy in heavy-fermion compounds
}}

\author{Ewan Scott}
\affiliation{Department of Mathematics, University College London, Gordon St., London WC1H 0AY, United Kingdom}

\author{Michal Kwasigroch}
\affiliation{Department of Mathematics, University College London, Gordon St., London WC1H 0AY, United Kingdom}
\affiliation{Trinity College, Cambridge, CB2 1TQ, United Kingdom}
\date{\today}

\begin{abstract}
\textcolor{blue}{
The local magnetic anisotropy of a typical crystalline compound is usually attributed to the combined effect of crystal electric fields and spin-orbit coupling. We show that this simple local picture is transformed in heavy-fermion compounds by the development of coherent electron scattering from local spin degrees of freedom. Provided the dominance of the coherence energy scale over the magnetic energy scale is strong enough, the fractionalisation and delocalisation of the spins destabilises their single-ion anisotropy by generating an opposing anisotropy in the exchange. Experimentally, this can manifest as competing splittings in the Curie-Weiss constants and effective moments. We show that in the presence of orthorhombic or tetragonal symmetry the destabilisation of the anisotropy can result in either ferromagnetic or antiferromagnetic order that is perpendicular to the high-temperature easy axis. In the absence of destabilisation, we show that the order is more likely to be antiferromagnetic. 
 In agreement with our theory, we also observe that the temperature at which the anisotropy of the uniform magnetic response changes tracks the coherence energy scale in a wide range of actinide and lanthanide compounds.
}
\end{abstract}

\maketitle

{\it Introduction.} -- 
Heavy-fermion compounds entangle electronic and local magnetic degrees of freedom leading to a rich variety of unusual magnetic phenomena such as  high-field superconductivity mediated by spin fluctuations \cite{Nature_UTE2_PHASEDia,Lewin_2023, Aoki_2022,Aoki_U_review}, with highly anisotropic and multiple superconducting phases, or metamagnetic jumps in the magnetisation \cite{doi:10.7566/JPSJ.88.063706}.
The interweaving begins with the formation of the flat heavy-fermion bands as the scattering of electrons from local magnetic moments becomes coherent. The heavy Fermi liquid excitations that emerge coexist with the residual local moments that form at the same time. Although many general questions remain, this `two-fluid' picture is a useful perspective in which to view a number of experimental observations \cite{Pines_two, Lonzarich_review, Yang_review,UAs_2_coexistence}, as well as theoretical results \cite{Kotliar_coexistence, Assad_Monte_Carlo, Perkins_Lacroix}. 

The rich phenomenology of magnetism in heavy-fermion metals is generally sensitive to the underlying crystal structure, crystal electric fields (CEF), spin-orbit coupling, or valency of the local-moment ions. Nonetheless, across a wide range of lanthanide and actinide compounds some robust universal features emerge \cite{Brando, Aoki_U_review, Eu_review}. One example is the reorientation of the magnetic easy axis with temperature, that is strongly correlated with the presence of the coherence maximum in the material's resistivity.
The resistivity rises logarithmically under cooling, and reaches a maximum at a characteristic temperature $T^{\ast}_{\rm coh}$, before dropping dramatically as the Kondo scattering of electrons from local spin degrees of freedom becomes coherent (see the bottom panel of Fig. 1). 
\textcolor{blue}
{Many heavy-fermion compounds with orthorhombic or tetragonal symmetry that order magnetically below this coherence temperature, do so at 90 degrees to the preferred direction of the local moments at high temperatures. The easy axis in the ordered state is thus perpendicular to the easy axis at temperatures above the coherence and exchange energy scales, where it is set by the single-ion anisotropy. In this Letter, we show that it is the development of coherence and flat heavy-fermion bands that destablises the local (single-ion) anisotropy and is responsible for the reorientation of the magnetic easy axis.}

\textcolor{blue}{
Before the onset of magnetic order, the destabilisation of the single-ion anisotropy can also be signaled by a change in the anisotropy of the uniform magnetic response, which is seen as a crossing of uniform magnetic susceptibilities at a temperature $T_{\rm cross}$  for fields along two perpendicular directions (see the bottom panel of Fig. 1).}
The connection between $T^{\ast}_{\rm coh}$ and $T_{\rm cross}$ was first noticed across a wide range of Yb and Ce based Kondo ferromagnets \cite{Brando}, where the \textcolor{blue}{susceptibilities cross} just before the onset of magnetic order and the crossing is preceded by a coherence maximum at a higher temperature. Here, we widen the universality of this phenomenon by including Eu and U based heavy-fermion \textcolor{blue}{ferromagnets, antiferromagnets, and paramagnets.}
\begin{figure}[h]
    \centering
    \includegraphics[width=0.48\textwidth]{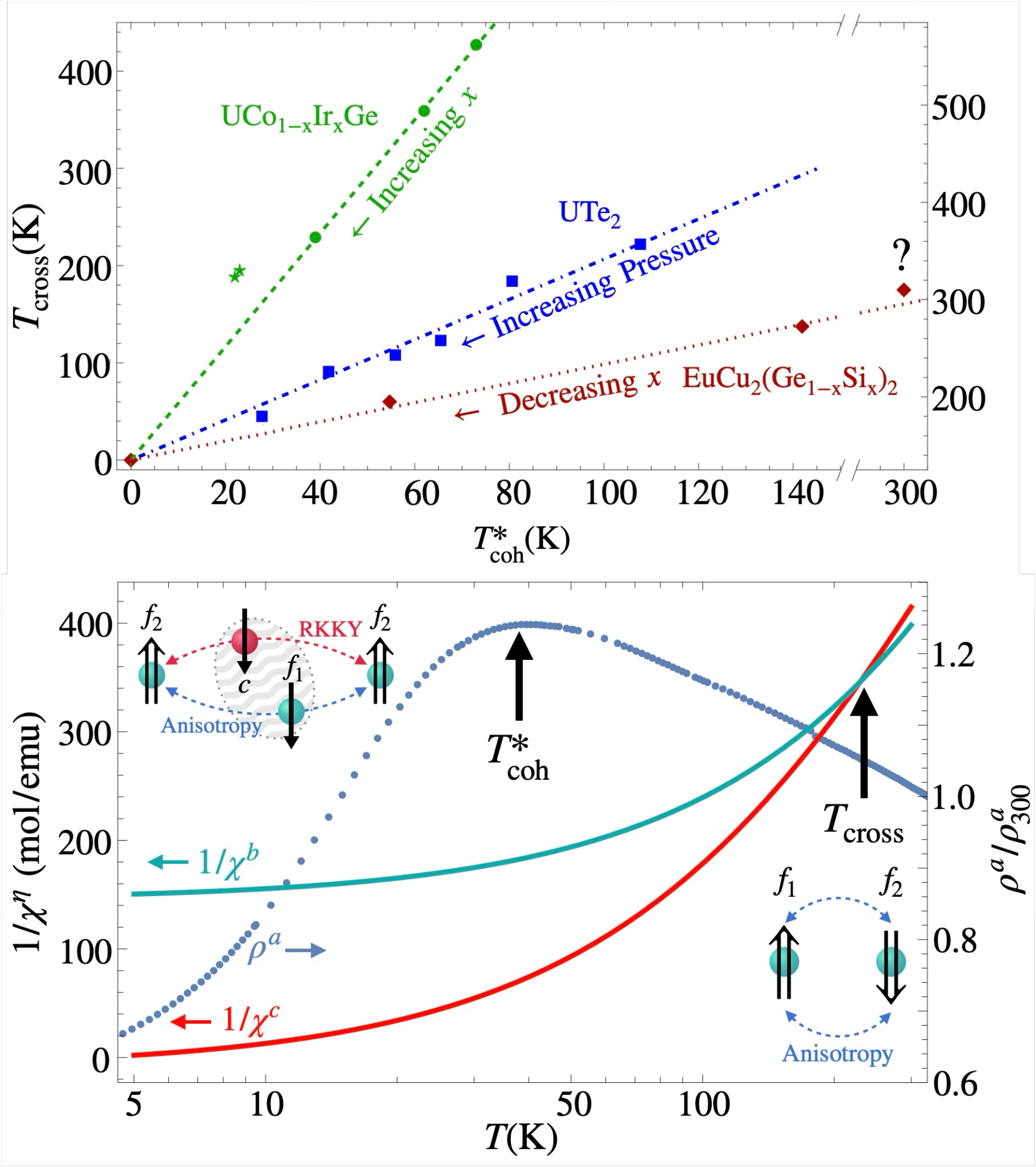}
    \caption{Top: The susceptibility crossing temperature, $T_{\rm cross}$, against the coherence temperature 
    $T^*_{\rm coh}$, for three heavy-fermion compounds.
    The {\it direct proportionality} fits suggest that the same energy scale is responsible for the development of coherence and easy-axis reorientation. (For UCo$_{1-x}$Ir$_{x}$Ge, the fit does not include the stars, and for EuCu$_2$(Ge$_{1-x}$Si$_x$)$_2$ it does not include the highest $T_{\rm cross}$, which is a tentative prediction.) Bottom: A plot of the $a$-axis resistivity and modified Curie-Weiss fits of $b$- and $c$-axis susceptibilities for UCo$_{0.98}$Ir$_{0.02}$Ge identifying $T^*_{\rm coh}$ and $T_{\rm cross}$. \textcolor{blue}{A single-ion anisotropy, that favours antialignment of the two components of the spin-1 moment along the $c$-axis, leads to a lower paramagnetic response along this direction at high temperature. Around $T^{*}_{\rm coh}$, when one of the components delocalises and becomes part of the heavy-fermion sea, the anisotropy interaction becomes non-local and a stronger RKKY exchange along the c-axis is generated.}
   [A detailed description of all the plotted data and any extrapolations used can be found in Table I in the End Matter.]}
    \label{fig:enter-label}
\end{figure}
Unlike in Ce or Yb based materials, in Eu or U based heavy-fermion compounds, \textcolor{blue}{the crossing of the susceptibilities} can take place far away from the onset of magnetic order, or even in its absence, \textcolor{blue}{showing that it is unlikely to be caused by the collective effects associated with the onset of magnetic ordering.} 
Remarkably, we find that, in these compounds, the crossing and coherence temperatures appear to be {\it directly proportional}, showing that the same energy scale is responsible for both (see Fig.\ 1 (top), with a detailed description of all the plotted data given in Table I in the End Matter).  

\textcolor{blue}{We find that single-ion anisotropy is most likely to be destabilised in heavy-fermion compounds that display more itinerant rather than localised magnetism. They are characterised by clear resistivity maxima at a temperature $T^*_{\rm coh}$ that is significantly higher than the temperature at which potential magnetic order sets in. As shown in Table I in the End Matter, they also often possess higher Sommerfeld coefficients and smaller ordered moments, again pointing to a greater degree of itinerancy. We show that ferromagnetic order and antiferromagnetic  order are both possible along the high-temperature hard axis, although the latter relies on Fermi surface nesting and is therefore more specific in nature, and perhaps less commonly seen. In the absence of destabilisation, antiferromagnetic order appears to be more common.}


In some compounds, \textcolor{blue}{e.g. URhGe,} coherence develops at a temperature $T^{\ast}_{\rm coh}$ that is much higher than the temperature at which \textcolor{blue}{fairly large moments order}, pointing to the underscreened nature of the underlying Kondo lattice, where in spite of the Kondo energy scale $T_K$ being much higher than the RKKY energy scale $T_{\rm RKKY}$, there are not enough conduction electrons to fractionalise the local spin degrees of freedom entirely and a so-called residual moment remains.
Motivated by the above observation, we present a generalisation of the standard large-$N$ Read-Newns theory to the underscreened Kondo lattice.
We will focus on the $S=1$ case.


{\it Model.} -- The underscreened $S=1$ Kondo lattice with a generic single-ion anisotropy, that is endowed to the spins through the combined effect of CEF and spin-orbit coupling, is described by the following Hamiltonian
\begin{equation}\label{Hamil} 
\begin{aligned}
H=&J_K\sum_{i}\mathbf{S}(\mathbf{r}_i)\cdot\mathbf{s}(\mathbf{r}_i)+\sum_{\mathbf{k}\sigma}
\epsilon_\mathbf{k} c^\dagger_{\sigma}(\mathbf{k})c_{\sigma}(\mathbf{k})\\ &+\sum_{\eta i} D^{\eta}\left(S_{}^{\eta}(\mathbf{r}_i)\right)^2
\textcolor{blue}{-\sum_{i} \mathbf{h}_i\cdot\left( \mathbf{S}(\mathbf{r}_i)+ \mathbf{s}(\mathbf{r}_i) \right)} ,
\end{aligned}
\end{equation}
where $J_K>0$ is the Kondo coupling between local spin $\mathbf{S}_i$ and the conduction electron spin $\mathbf{s}_i$, $\sigma$ denote the up/down spin indices, $c_{\sigma}(\mathbf{k})$ are the usual conduction electron annihilation operators, \textcolor{blue}{and $\mathbf{h}$ is an external magnetic field.}  For simplicity, we have assumed that electrons and spins occupy the same Bravais lattice sites, and $\epsilon_{\mathbf{k}}$ is the dispersion of the single conduction band \textcolor{blue}{with bandwidth $W$.} \textcolor{blue}{Unless stated otherwise, we will be using a constant density of states $\rho=1/W$.} $D^{x,y,z}$ is a general orthorhombic single-ion anisotropy for spin-1, \textcolor{blue}{and we have set its average value to zero.} We will work away from the insulating limit, and unless specified otherwise, just below half-filling.

We write the $S=1$ local moments in the Schwinger fermion representation \cite{PhysRevB.76.125101, PhysRevB.87.205107},
\begin{equation}\label{abrikosov}
S^{\eta}(\mathbf{r}_i)= \sum_{a=1,2} S^{\eta}_{a}(\mathbf{r}_i) =
\frac{1}{2}\sum_{a\sigma\sigma'} f_{a\sigma}^\dagger(\mathbf{r}_i) \sigma^{\eta}_{\sigma\sigma'} f_{a\sigma'}(\mathbf{r}_i),
\end{equation}
where $a=1,2$ index the two spin-1/2 fermions, $\sigma, \sigma'$ denote the up/down spin indices,  $\sigma^{\eta}$ are Pauli matrices with $\eta\in \{x,y,z\}$, and $f_{a\sigma}(\mathbf{r}_i)$ obey the usual fermionic anticommutation relations. We will also promote the $SU(2)$ spin degrees of freedom to $SU(N)$. This can be done in a number of ways and we will choose to have $N/2$ replicas of spin-up states and an equal number of spin-down states on each lattice site, with half of them occupied, so that we are connected most straightforwardly to the physical $N=2$ limit. $N$ will be the control parameter for the large-$N$ theory. The Hamiltonian can be written (up to a constant) as 
\begin{equation}\label{FermiHam}
\begin{aligned}
H_N=&\frac{J_K}{N}\sum_{a\alpha\sigma\beta\sigma' i}f^\dagger_{a\alpha\sigma }(\mathbf{r}_i)f_{a\beta\sigma' }(\mathbf{r}_i) c^\dagger_{\beta \sigma' }(\mathbf{r}_i)c_{\alpha\sigma }(\mathbf{r}_i)
\nonumber\\
&+\sum_{\mathbf{k}\alpha \sigma}\epsilon_\mathbf{k} c^\dagger_{\alpha\sigma}(\mathbf{k})c_{\alpha\sigma}(\mathbf{k}) 
+ \sum_{i \alpha \eta} D^{\eta} \left( \sum_a S^{\eta}_{a\alpha}(\mathbf{r}_i)\right)^2 \\ &\textcolor{blue}{-\sum_{i\eta\alpha } h_i^\eta\left( \sum_{a} S^\eta_{a\alpha}(\mathbf{r}_i)+ s^\eta_{\alpha}(\mathbf{r}_i) \right)},
\end{aligned}
\end{equation}
where
\begin{eqnarray}
   S^{\eta}_{a \alpha}(\mathbf{r}_i)= \frac{1}{2}\sum_{\sigma \sigma'} f^{\dagger}_{a\alpha\sigma}(\mathbf{r}_i) \sigma^{\eta}_{\sigma \sigma'}f_{a\alpha \sigma'}(\mathbf{r}_i), 
\end{eqnarray}
and $\alpha, \beta$ index the spin up/down replicas. 
 The enlargement of the Hilbert space through the Schwinger fermion representation comes with several constraints and associated gauge symmetries:
\begin{eqnarray}\label{constraint}
    \hat{n}^f_i&:=& \sum_{a=1,2} \hat{n}^f_{a i} = \sum_{a\alpha\sigma} f^{\dagger}_{a\alpha\sigma}(\mathbf{r}_i) f_{a\alpha\sigma}(\mathbf{r}_i) = N,
    \nonumber\\
   \hat{{T}}^{\eta}_i &:=& \frac{1}{2}\sum_{a a' \alpha\sigma} f^{\dagger}_{a\alpha\sigma}(\mathbf{r}_i) \tau^{\eta}_{a a'} f_{a'\alpha\sigma}(\mathbf{r}_i) = 0,
\end{eqnarray}
where $\tau^{x,y,z}$ are Pauli matrices, that now act on the 2-dimensional $a$ subspace. We can understand the constraints on $\hat{n}^f_i$ and $\hat{T}^z_i$ as together ensuring that there are $\hat{n}^f_{1 i}=\frac{N}{2}$ $f_{1}$-fermions and  $\hat{n}^f_{2 i}=\frac{N}{2}$ $f_{2}$-fermions on each lattice site. The constraints $\hat{T}^{x,y}$ couple the $SU(N)$ $f_1$ and $f_2$ spins, which for $N=2$ projects out the singlet state and corresponds to an infinite Hund's coupling between $f_1$ and $f_2$ moments. The constraints generate the  $U(2) \sim U(1) \times SU(2) $ group of local gauge transformations.

\textcolor{blue}{Read-Newns large-$N$ theory introduces the hybridisation field
 \begin{equation}
     \begin{pmatrix}
         \hat{V}_{1i}\\ \hat{V}_{2i}
     \end{pmatrix}=\frac{1}{N}\sum_{\alpha\sigma} 
     \begin{pmatrix}
          f^\dagger_{1\alpha\sigma}(\mathbf{r}_i)c_{\alpha\sigma}(\mathbf{r}_i)\\
         f^\dagger_{2\alpha\sigma}(\mathbf{r}_i)c_{\alpha\sigma}(\mathbf{r}_i)
     \end{pmatrix}.
 \end{equation}
In the saddle-point approximation, the local $U(2)$ gauge symmetry can be used to set $\langle\hat{V}_{2i}\rangle=0$ and $\langle \hat{V}_{1i} \rangle >0$. $f_2$-moments thus decouple from conduction electrons at the level of large-N mean-field theory \cite{PhysRevB.87.205107}}. 

{\it Residual local moments.} --
\textcolor{blue}{$\mathcal{O}(1/N)$ corrections to the saddle-point approximation break the macroscopic degeneracy of the $f_2$-moment subspace \cite{PhysRevB.87.205107}.
This can be captured with the following variational ansatz for the low-temperature state }
\begin{equation}
    \label{ansatz}
    | \Psi\rangle = \prod_{\mathbf{k\alpha\sigma \nu}} \left(u^{\nu}_{\mathbf{k}\sigma} c^{\dagger}_{\alpha\sigma}(\mathbf{k})+ v^{\nu}_{\mathbf{k}\sigma}f^{\dagger}_{1\alpha\sigma}(\mathbf{k})\right) |0\rangle \otimes  \prod_{\alpha i} | \psi_{2\alpha i} \rangle,
\end{equation}
where $\nu$ indexes two heavy-fermion bands and $|\psi_{2\alpha i} \rangle$ is the state of one of the $f_2$-moments. By employing the usual static approximation for the $f_2$-moments, calculating $\langle \Psi | H_N | \Psi \rangle$ with $\{u^\nu_{\mathbf{k} \sigma}, v^\nu_{\mathbf{k}  \sigma}\}$ chosen to minimise the expectation value subject to global constraints on the number of conduction and $f_1$-fermions, we can derive the effective RKKY interaction between $f_2$-moments that is mediated by the heavy fermions. Expanding to quadratic order in $1/N$ and $D^\eta$ and linear order in an external magnetic field $h$, we obtain
\begin{equation}\label{Heff}
     H_{\rm eff}\! =\!-\frac{1}{2}\textcolor{blue}{\sum_{i\neq  j ,\eta}  \mathcal{J}^{\eta} (\mathbf{r}_i-\mathbf{r}_j) S^{\eta}_{2}(\mathbf{r}_i) S^{\eta}_{2}(\mathbf{r}_j)}\! -\! \sum_{\mathbf{q} \eta}\tilde{h}^{\eta}(\mathbf{q})
     S^{\eta}_2(\mathbf{q}),
\end{equation}
\textcolor{blue}{where $S_2^{\eta}(\mathbf{q}) = \frac{1}{\sqrt{N_s}}\sum_i S^{\eta}_2(\mathbf{r}_i)e^{-i\mathbf{q}\cdot\mathbf{r}_i}$, $\mathcal{J}^{\eta}(\mathbf{q})=\sum_{i} \mathcal{J}^{\eta}(\mathbf{r}_i)e^{-i\mathbf{q}\cdot\mathbf{r}_i}$, $N_s$ is the number of lattice sites, and}
\begin{align}
   & \mathcal{J}^{\eta} (\mathbf{q}) =  \frac{4J_K^2}{N^2} \chi_{cc}(\mathbf{q})+ \frac{8J_K}{N}D^{\eta} \chi_{cf} (\mathbf{q}) +(2D^{\eta})^2 \chi_{ff}(\mathbf{q}),
    \nonumber\\
    &\frac{\tilde{h}(\mathbf{q})}{h(\mathbf{q})} \! =\! 1\!-\! \frac{2J_K}{N}(\chi_{cc}(\mathbf{q})+\chi_{cf}(\mathbf{q}))\\&\quad\quad\quad\quad\quad\quad\quad\quad\quad\quad -2D^{\eta}
    \left(
    \chi_{ff}(\mathbf{q}) + \chi_{cf}(\mathbf{q})
    \right),\nonumber
    \label{eq: RKKY reinforcement}
\end{align}
and we have dropped the replica index in $H_{\rm eff}$, reverting to the physical case $N=2$, so that $S^{\eta}_2(\mathbf{r}_i)$ are $SU(2)$ spin operators. 
The susceptibilities of the $N\rightarrow\infty$ ground state ($f_2$-moments completely decouple in this limit) are given by
\begin{eqnarray}
    \chi_{cc}(\mathbf{q}) &=& \frac{1}{2N_s} \sum_{\mathbf{k}\nu\nu'} |u^{\nu}_{\mathbf{k}}|^2 |u^{\nu'}_{\mathbf{k}+\mathbf{q}}|^2
      \frac{n(E^{\nu'}_{\mathbf{k}+\mathbf{q}})-n(E^{\nu}_{\mathbf{k}})}{E^{\nu}_{\mathbf{k}}-E^{\nu'}_{\mathbf{k}+\mathbf{q}}},
      \nonumber\\
       \chi_{cf}(\mathbf{q}) &=& \frac{1}{2N_s} \sum_{\mathbf{k}\nu\nu'} \bar{v}^{\nu}_{\mathbf{k}}v^{\nu'}_{\mathbf{k}+\mathbf{q}} u^{\nu}_{\mathbf{k}}\bar{u}^{\nu'}_{\mathbf{k}+\mathbf{q}}
      \frac{n(E^{\nu'}_{\mathbf{k}+\mathbf{q}})-n(E^{\nu}_{\mathbf{k}})}{E^{\nu}_{\mathbf{k}}-E^{\nu'}_{\mathbf{k}+\mathbf{q}}},
      \nonumber\\
       \chi_{ff}(\mathbf{q}) &=& \frac{1}{2N_s} \sum_{\mathbf{k}\nu\nu'} |v^{\nu}_{\mathbf{k}}|^2 |v^{\nu'}_{\mathbf{k}+\mathbf{q}}|^2
      \frac{n(E^{\nu'}_{\mathbf{k}+\mathbf{q}})-n(E^{\nu}_{\mathbf{k}})}{E^{\nu}_{\mathbf{k}}-E^{\nu'}_{\mathbf{k}+\mathbf{q}}},
\end{eqnarray}
where $\chi_{cc}(\mathbf{q})$ ($\chi_{cf}(\mathbf{q})$) gives the magnetic susceptibility of the conduction electrons ($f_1$ local moments) when a $\mathbf{q}$-modulated magnetic field is applied isothermally to conduction electrons only. Similarly, $\chi_{ff}(\mathbf{q})$ gives the magnetic susceptibility of the $f_1$ local moments, when the applied field only couples to them. $E^\nu_{\mathbf{k}}$ is the energy dispersion of the heavy-fermion quasiparticles in band $\nu$ and $n(E^\nu_{\mathbf{k}})$ the corresponding Fermi occupation number. (Expressions for $u^\nu_{\mathbf{k} }, v^\nu_{\mathbf{k}  }$ and $E^\nu_{\mathbf{k}}$ \textcolor{blue}{and an outline of the derivation of the effective Hamiltonian} can be found in the End Matter.) In the isotropic limit, our effective Hamiltonian agrees with the one derived in Ref. \cite{PhysRevB.87.205107} for an underscreened spin$-1$ Kondo chain.


\begin{figure}\label{fig:SC:WC}
    \centering
    \includegraphics[width=0.48\textwidth]{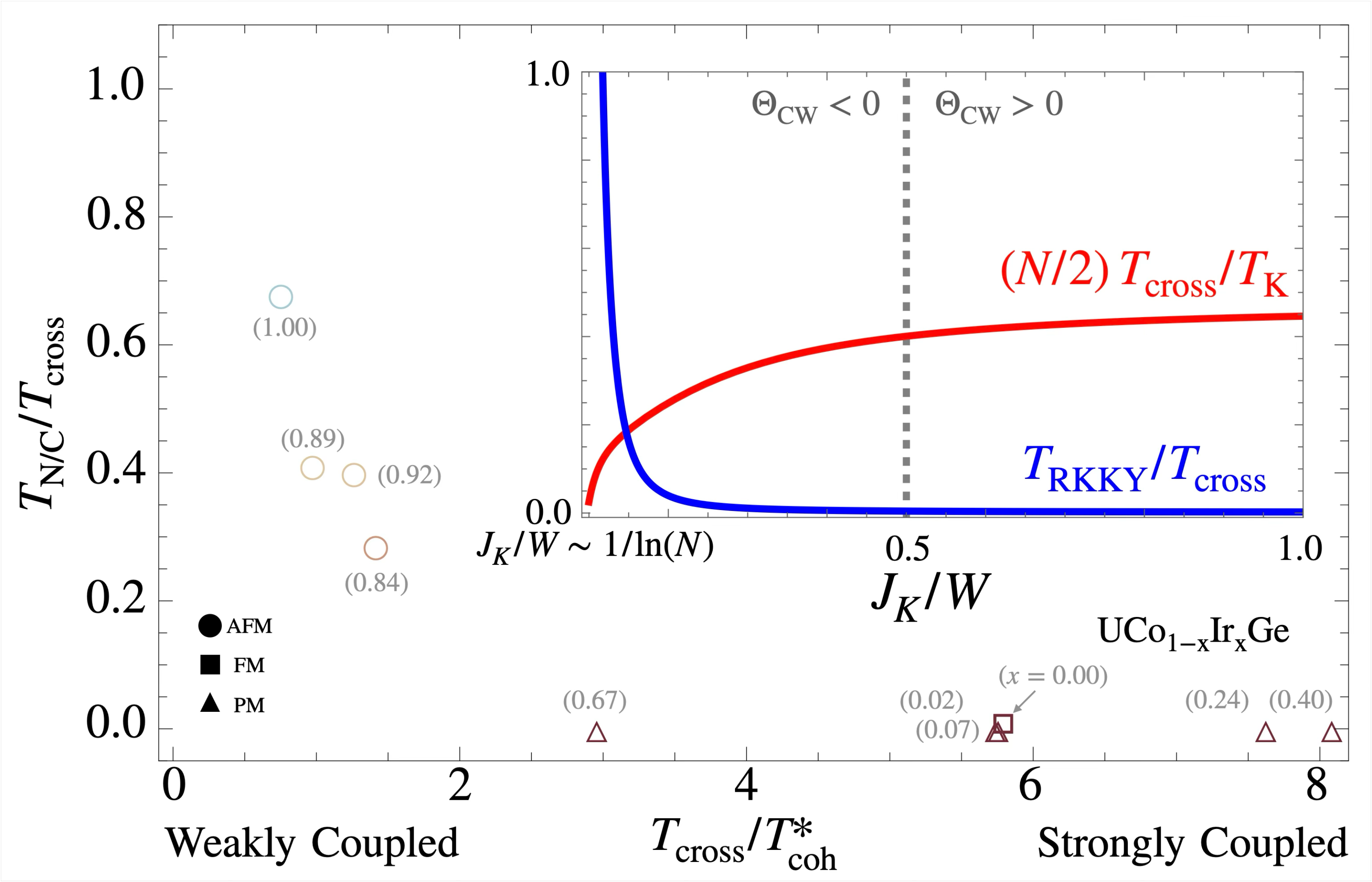}
    \caption{The ratios $T_{\rm N/C}/T_{\rm cross}$ and $T_{\rm cross}/T^*_{\rm coh}$ plotted for UCo$_{1-x}$Ir$_{x}$Ge, where $T_{\rm cross}$ is the uniform susceptibility crossing temperature, $T^*_{\rm coh}$ the coherence temperature, and $T_{\rm N/C}$ the \textcolor{blue}{Neel (circles) or Curie (squares) temperature depending on the type of magnetic order. Triangles denote lack of magnetic order.}  \textcolor{blue}{Inset: the ratios $T_{\rm RKKY}/T_{\rm cross}$ and $(N/2)T_{\rm cross}/T_K$ plotted against the Kondo coupling strength, $J_K/W$. There are two regimes: weak coupling at $J_K/W \lesssim 1/2$, where the Curie-Weiss constant is negative and  $T_{\rm cross}$ approaches $T_{\rm RKKY}$ before vanishing entirely around $J_K/W \sim 1/\ln{N}$, and strong coupling at $J_K/W \gtrsim 1/2$, where the Curie-Weiss constant is positive and $ T_{\rm RKKY}\ll T_{\rm cross}\approx T_K/N $.}}
\end{figure}

{\it Easy-axis reorientation.} -- \textcolor{blue}{Clearly at temperatures much greater than $T_K$ and $T_{\rm RKKY}$, where the spin-1 moments are unhybridised and RKKY interactions can be neglected, a hard-axis anisotropy $D^{\eta}>0$ along $\eta$ leads to a smaller magnetic response along this axis. Let us now consider a magnetic modulation with wavevector $\mathbf{q}$ at temperatures below $T_K$. Assuming for now (see later discussion) that $\chi(\mathbf{q})$ and $\chi(\mathbf{q})-\bar{\chi}$ are all positive for $T \ll T_K$ (the bar denotes an average of the respective function over the Brillouin zone)}, the anisotropy of the effective field, \textcolor{blue}{$\tilde{h}^{\eta}(\mathbf{q})$}, and the anisotropy of the RKKY exchange \textcolor{blue}{$\mathcal{J}^{\eta}(\mathbf{q})-\bar{\mathcal{J}}^{\eta}$} are of opposite sign to first order in $D^{\eta}$. For example, a hard-axis anisotropy $D^{\eta}>0$ along $\eta$ leads to a reduction of the residual effective moment but a reinforcement of RKKY exchange along this direction. We can understand this by looking at how the residual moment polarises the surrounding heavy Fermi sea. The greater the $D^{\eta}$ the stronger the local anti-alignment induced by the anisotropy interaction between the $f_2$, and the now delocalised, $f_1$-moments (see the bottom panel of Fig. 1). Locally this gives a lower effective moment along the direction with higher $D^{\eta}$, and therefore lower paramagnetic response at high temperature, but globally, the long-range RKKY forces are stronger along this direction. The collective fluctuations generated by RKKY dominate at low temperatures and the higher the $D^{\eta}$ the stronger the magnetic response.
In particular, \textcolor{blue}{a magnetically ordered} collinear ground state would have moments along the direction with the highest $D^{\eta}$.



The local and exchange anisotropies respectively correspond to the anisotropy in the effective moments $\mu_{\rm eff}^{\eta}$ and Curie-Weiss constants $\Theta_{\rm CW}^{\eta}$. Indeed, in agreement with $H_{\rm eff}$, many of the compounds surveyed have $(\Delta \mu_{\rm eff}^{\eta})/\mu_{\rm eff} \sim - (\Delta \Theta_{\rm CW}^{\eta})/T_{\rm cross}$, where $\Delta$ takes the difference between the two directions involved in uniform susceptibility crossing and $\mu_{\rm eff}$ is the average effective moment for these two directions. The competition between the two anisotropies leads to a crossing of the uniform susceptibilities at a temperature of\textcolor{blue}{
\begin{eqnarray}
    &&T_{\rm cross}= \frac{J^2_K}{N^2}\left(\chi_{cc}(\mathbf{0}) -\bar{\chi}_{cc}\right)+
 \nonumber\\
    &&\frac{J_K}{N}  \frac{\chi_{cf}(\mathbf{0})- \bar{\chi}_{cf}}{\chi_{ff}(\mathbf{0})+ \chi_{cf}(\mathbf{0})} \left( 1 -\frac{2J_K}{N} \left(\chi_{cc}(\mathbf{0}) + \chi_{cf}(\mathbf{0}) \right) \right),
      \nonumber\\
\end{eqnarray}}
\textcolor{blue}{in the Weiss mean-field approximation,} and to quadratic order in $1/N$ and first non-vanishing order in $D^{\eta}$. 

\textcolor{blue}{We identify two regimes. In the strong-coupling limit $J_K/W \gtrsim 1/2$, we have to first non-vanishing order in $D^{\eta}$ and $1/N$
\begin{eqnarray}
    T_{\rm cross} \approx \frac{T_K}{N},
\end{eqnarray}
where $T_K$ is the Kondo energy scale} \footnote{We take it to be equal to the difference between the energy of the $N\rightarrow\infty$ ground state and the $N\rightarrow\infty, J_K=0$ Fermi gas state.}. 
\textcolor{blue}{Extrapolating to the physical case $N=2$, we find that the crossing temperature is proportional to the strength of the Kondo interaction, i.e., we should have $T_{\rm cross} \propto T^*_{\rm coh}$, which indeed shows excellent agreement with experimental data in a wide range of heavy-fermion compounds as shown in Fig.\ 1.}

\textcolor{blue}{In the weak-coupling limit, for any fixed $N$, as $J_K$ is reduced, the crossing temperature approaches the RKKY energy scale, which we define as $T_{\rm RKKY} = |\Theta_{\rm CW}|=(J_K/N)^2 |\chi_{cc}(\mathbf{0})-\bar{\chi}_{cc}|$, before vanishing entirely. We also find that the sign of the $D^{\eta}=0$ Curie-Weiss constant $\Theta_{\rm CW}$ changes from positive to negative around $J/W\sim 1/2$ as we transition from the strong to weak coupling regime, suggesting that the ground state is ferromagnetic at strong-coupling and antiferromagnetic at weak-coupling. The fact that we see this for a constant density of states approximation, where there is no specific underlying lattice, shows that this is a universal phenomenon, which we indeed observe for other lattices (see our later discussion).} Fig.\ 2 illustrates the two regimes for both theory and for experimentally measured values for UCo$_{1-x}$Ir$_{x}$Ge, which  spans the range from strongly to weakly coupled as a function of the doping $x$.

\begin{figure}
    \centering
    \includegraphics[width=0.48\textwidth]{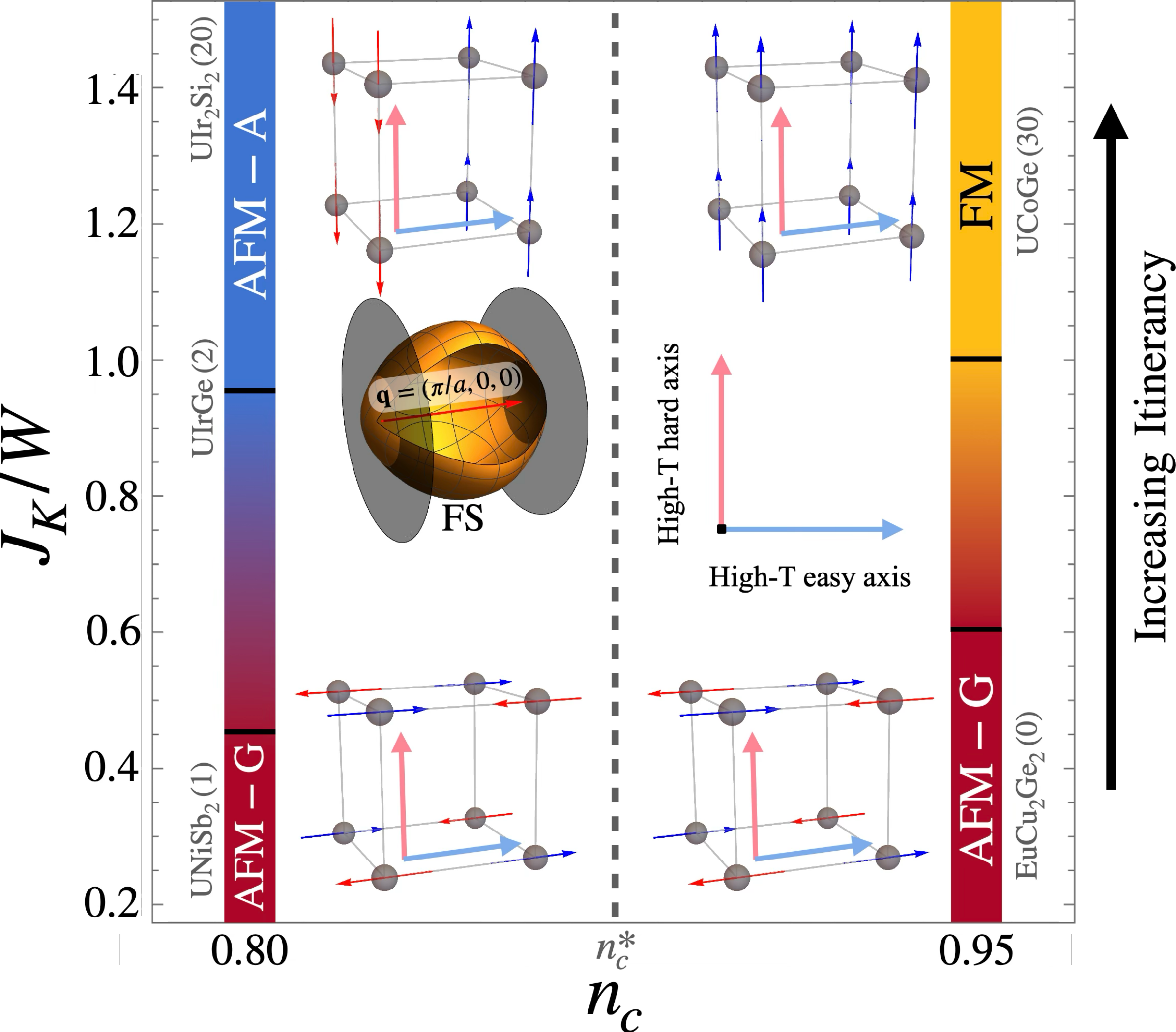}
    \label{fig:phase_diag}
    \caption{\textcolor{blue}{The ground state phase diagram of the $S=1$ simple cubic Kondo lattice in the weak anisotropy limit $D^{\eta}\rightarrow 0$ for two representative values of the electron filling $n_c$ as a function of the Kondo coupling $J_K$ normalised by the bandwidth $W$. There is an AFM-G phase with spins along the high-temperature easy axis at low coupling and an AFM-A or FM phase with spins along the high-temperature hard axis at large $J_K$. For each phase, we give a representative compound from Table I in the End Matter and state the ratio of $T^{\star}_{\rm coh.}/T_{N/C}$, where $T^{\star}_{\rm coh.}$ is the coherence temperature and $T_{N/C}$ the Neel or Curie temperature. Higher ratios signal more itinerant magnetism. The AFM-A phase relies on nesting. The Fermi surface and the locus of points which are connected by the AFM wavevector to a point of equal energy have been shown. $n_c^*$ denotes the region where the crossing from AFM-A to FM occurs at large $J_K$. The gaps between phases indicate regions where energy differences between competing phases were smaller than the numerical error and/or there is non-collinear magnetic order.}}
\end{figure}

\textcolor{blue}{{\it   Magnetic ground states} -- We apply our effective Hamiltonian to the representative case of the simple cubic lattice with nearest neighbour hopping (similar behaviour observed for other orthorhombic lattices) and determine its classical $T=0$ phases in the weak anisotropy limit $D^{\eta}\rightarrow 0$ for a range of conduction electron lattice fillings $0.75<n_c<1$ and coupling constants $1/6<J_K/W<3$. To simplify the discussion of easy-axis reorientation, we restrict ourselves to simple collinear phases, where the location of the maximum of $\chi_{cc}(\mathbf{q})$ determines the magnetic ordering wavevector $\mathbf{q}$ and if $\chi_{cf}(\mathbf{q})-\bar{\chi}_{cf}$ is positive (negative) at this wavevector the magnetic order is along the high-temperature hard (easy) axis.}

\textcolor{blue}{Our results are shown in Fig.\ 3 for two representative fillings. We can see that order takes place along the high-temperature hard-axis at strong enough Kondo coupling, when the excitation energy for interband particle-hole fluctuations is sufficiently large and intraband fluctuations dominate. $\chi_{cf}(\mathbf{q})$ is manifestly positive in this limit (and $\chi_{cf}(\mathbf{q})-\bar{\chi}_{cf}$ at the ordering wavevector if $\chi_{cf}(\mathbf{q})$ and $\chi_{cc}(\mathbf{q})$ are maximised at similar wavevectors). Stronger Kondo coupling (at fixed $N\gg 1$ in our theory) corresponds to a greater dominance of the Kondo energy scale over the RKKY energy scale and thus more itinerant magnetism. We can indeed see that compounds with higher ratios of $T^*_{\rm coh}/T_{N,C}$ are more likely to exhibit order along the high-temperature hard-axis. While ferromagnetic and antiferromagnetic magnetic order both can take place, the latter is favoured when a finite momentum transfer connects Fermi surface points of the same energy, and is thus lattice and filling specific
by nature, and perhaps therefore less common.}

\textcolor{blue}{At weak coupling interband particle-hole fluctuations dominate. They enforce the antiferromagnetic correlation between conduction and $f_1$-electrons, and $\chi_{cf}(\mathbf{q})<0$ (and $\chi_{cf}(\mathbf{q})-\bar{\chi}_{cf}<0$ at the ordering wavevector if $\chi_{cf}(\mathbf{q})$ and $\chi_{cc}(\mathbf{q})$ are extremised at similar wavevectors). Because the indirect band gap is smaller than the direct one, the antiferromagnetic $\mathbf{q}=(\pi/a,\pi/a,\pi/a)$ phase also dominates in this limit.   }




{\it Conclusion} -- In this Letter, we have shown that the development of flat bands in heavy-fermion metals \textcolor{blue}{can destabilise the single-ion magnetic anisotropy of the local moments, that sets the easy axis at high temperatures, by generating an opposing exchange anisotropy. The destabilisation leads to a reorientation of the easy-axis with cooling and magnetic order (if it takes place) at 90 degrees to the high-temperature easy-axis. The destabilisation is mostly likely to occur in more itinerant heavy-fermion compounds that have a clear resistivity maximum at a temperature significantly higher than the magnetic ordering temperature.} 

{\it Acknowledgments} -- We acknowledge the following people for providing experimental data on ${\rm UCo}_{1-x}{\rm Ir}_{x}{\rm Ge}$ and ${\rm URh}_{1-x}{\rm Ir}_{x}{\rm Ge}$: V. Sechovský, D. Hovančík, and J. Pospíšil and F. Malte Grosche, A. Eaton, Z. Wu for providing experimental data on ${\rm UTe}_2$. As well as those already mentioned, we would like to thank the following people for stimulating discussions: M. Vojta, M. Brando, C. Batista, P. Dai, M. Zhitomirsky, P. Coleman, D. Aoki, J. Annett, D. Adroja, F. Krüger, H. Hu, K. Wojcik, and G. Lonzarich for many imaginative ideas.

\bibliography{apssamp}

\clearpage
\appendix
\begin{table*}[]
\begin{tabular}{llllllllllll}
\hline
\hline
{Compound}           &{P }& $T^*_{\rm coh}$  & $T_{\rm cross}$  &$T_{\rm max}^{\rm hard}$  &$T_N$  & $T_C$ &\textcolor{blue}{Orien.} &\textcolor{blue}{HTEA} & \textcolor{blue}{$\mu\,$}&  \textcolor{blue}{$\gamma\,$} \\
          &{(GPa)}& (K)  &  (K)  & (K) & (K) &  (K)& & & $(\mu_{\rm B})$&  $({\rm mJmol^{-1}K^{-2}})$ \\\hline
${\rm UTe_2}$ (Ortho.)\cite{Akoi,Thomas_res,doi:10.7566/JPSJ.88.043702} &  0      & $108^{a\P}$      & 222                 &37 &      &      &NM&b&&120     \\
Crossing $b\rightarrow a$                                &  0.6                   & $81^{a\P}$            & 184                 &31 &      &      &NM&b&&--    \\
Metamagnetism observed                                   &  0.9                   & $65^{a\P}$            & 123                 &24 &      &      &NM&b&&--     \\
along b at amb. p.                                       &  1.1                   & $56^{a\P}$            & 108                 &22 &      &      &NM&b&&--     \\
                                                         &  1.4                   & $42^{a\P}$            & 91                  &16 &      &      &NM&b&&--     \\
                                                         &  1.7                   & $28^{a\P}$            & 45                  & & 3      &      &a?&b&$>0.3$&--     \\ \hline
                                                
{Compound}           &{$x$ }& $T^*_{\rm coh}$  & $T_{\rm cross}$  &$T_{\rm max}^{\rm hard}$  &$T_N$  & $T_C$ &\textcolor{blue}{Orien.} &\textcolor{blue}{HTEA} & \textcolor{blue}{$\mu\,$}&  \textcolor{blue}{$\gamma\,$} \\
          &{}& (K)  &  (K)  & (K) & (K) &  (K)& & & $(\mu_{\rm B})$&  $({\rm mJmol^{-1}K^{-2}})$ \\ \hline

${\rm UCo_{1-x}Ir_xGe}$ (Ortho.)\cite{PhysRevB.105.014436,doi:10.1080/14786431003630900,PROKES2004E199}& $0.00$ & $74^a$  & $427^{\dagger}$& &     & 2.5 &c&b&0.07&60      \\
Crossing $b\rightarrow c$                                                & 0.02   & $63^a$            & $359^{\dagger}$     &   &           &     &NM&b&& 55    \\
Metamagnetism observed                                                   & 0.07   & $40^a$            & $229$               &     &         &     &NM&b&& 40     \\
for $x>0.84$ along b                                                     & 0.24   & $24^a$            & $187$               &35   &         &     &NM&b&& 56     \\
                                                                         & 0.40   & $24^a$            & 194                 &30 &           &     &NM&b&& 65    \\
                                                                         & 0.67   & $15^a$            & 45                  &25 &           &     &NM&b&& 69     \\
                                                          & 0.84   & $25^a$     & 32                  &13& 10.2       &     &prin. c&b&--&    40  \\
                                                                         & 0.89   & $31^c$            & 27                  &18& 12.2       &     &prin. c&b&--&30   \\
                                                                         & 0.92   & $29^c$            & 36                  &19& 14.4       &     &prin. c&b&--&20      \\
                                                                         & 1.00   & $32^c$            & 21                  &29& 16.5       &     &prin. c&b&0.36&19      \\ \hline
${\rm URh_{1-x}Ir_xGe}$ (Ortho.)\cite{PhysRevB.95.155138,Aoki_2022,doi:10.7566/JPSJ.88.022001}                & 0.00   & $188^a$           & --                  & &  & 9.5 &c&NR&0.43&    163  \\
Crossing $b\rightarrow c$                                                & 0.14   & --                & 69$^\dagger$        & &             & 9.1 &c&b&0.39  & 160\\
Metamagnetism observed                                                   & 0.43   & --                & 34$^\dagger$        & &        & 6.2 &c&b&0.24   & 175   \\
along b                                                                  & 0.45   & --                & 45$^\dagger$        &$\sim 11$&     & 3.9 &c&b& --  & 120   \\
                                                                         & 0.58   & --                & 34$^\dagger$        &12& 7.0        &     &c&b&-- &   70   \\
                                                                         & 1.00   & $32^c$            & 21                  &29& 16.5       &     &prin. c&b&0.36&  19  \\ \hline
${\rm EuCu_2(Ge_{1-x}Si_x)_2}$ (Tetra.)\cite{doi:10.7566/JPSJ.87.064706,PhysRevB.94.195101,PhysRevB.69.014422} & $\leq0.6$    & $0^{a\ddagger}$& 0\textsuperscript{§}&& 18.8  &   & &\textcolor{blue}{NR}&$\geq 5.3$&--       \\
Crossing \textcolor{blue}{$a\rightarrow c$}                                                & 0.7    & $55^a$            & $\sim$ 60            & & 15.5       &     &c?&NR&$<5.3$&191 \\
                                                                         & 0.8    & $142^a$           & 138                 & &             &     &NM&a&&137       \\
                                                                         & 1.0    & $300^a$           & $\sim 320\,?$       &  $\sim 230$&  &     &NM&a&&25   \\ \hline                                                        
${\rm UIr_2Si_2}$ (Tetra.)\cite{UIr2Si2}                                 &        & $97^c$            & 60                  &6& 5.5         &     &c&a&0.1&292       \\
Crossing $a\rightarrow c$,                                               &        &                   &                     &  &            &     &&&&       \\ \hline
${\rm UPt_2Si_2}$ (Tetra.)\cite{BLECKMANN20102447,AMITSUKA1992173}       &        & $176^a$           & 76.5                &  & 32         &     &c&a&2.5&--       \\
Crossing $a\rightarrow c$                                                &        &                   &                     & &             &     &&&&       \\ 
Metamagnetism observed along a                                           &        &                   &                     & &             &     &&&&       \\ \hline
${\rm UIrSi_3}$ (Tetra.)\cite{PhysRevB.97.144423,PhysRevB.100.014401}    &        & $224^c$           & 93                  & & 41.7        &     &c&a&--&31       \\
Crossing $a\rightarrow c$                                                &        &                   &                     & &             &     &&&&       \\ \hline
${\rm EuNi_2P_2}$ (Tetra.)\cite{doi:10.7566/JPSJ.82.083708}              &        & $135^a$           & 76                  &46&            &     &NM&ab&&93      \\ 
Crossing $ab\rightarrow c$                                               &        &                   &                     &  &            &     &&&&      \\
Metamagnetism observed along ab                                          &        &                   &                     &  &            &     &&&&      \\ \hline
YbNi$_4$P$_2$ (Tetra.)\cite{Brando,Krellner_2011,doi:10.1126/science.1230583,YbNi4P2Gamma} &   &     $19^c$        &      0.16           &  &   &     0.15   &ab&c&$<0.05$&2000\\
Crossing $\parallel c\rightarrow \perp c$                                &        &                   &                     &  &            &     &&&&      \\ \hline
CeRuPO (Tetra.)\cite{Brando,KRELLNER20081875,PhysRevB.76.104418,PhysRevLett.100.066401} & &   $41^c$  &      $15^*$         &  &            &     15   &c& ab &0.3&--  \\
Crossing $\perp c\rightarrow \parallel c$                                &        &                   &                     & &             &     &&&&      \\ \hline
Yb(Rh$_{0.75}$Co$_{0.25}$)$_2$Si$_{2}$ (Tetra.)  & & $47^c$ & $1.3^*$ && & 1.3 &c&ab&0.1&850 \\
Crossing $\perp c\rightarrow \parallel c$\cite{Brando,PhysRevLett.110.256402,PhysRevB.83.144405,PhysRevB.85.035119,PhysRevB.83.144405}&&&      &  &            &      &&&&     \\ \hline
YbIr$_3$Ge$_7$ (Rhomb.)\cite{Brando,PhysRevB.99.121109}                  &        &     $7^c$         &      2.9     &  &            &     2.6   &ab&c&0.05&--  \\
Crossing $\parallel c\rightarrow \perp c$                                &        &                   &                     &  &            &      &&&&     \\
\hline
\textcolor{blue}{${\rm UGe_2}$ (Ortho.)}\cite{doi:10.7566/JPSJ.88.022001}   &        &      $70$        &   0\textsuperscript{§}               &    &          & 52      &a&a&1.48&34  \\
\hline
\textcolor{blue}{${\rm UNiSb_2}$ (Tetra.)}  \cite{2004_IKEDA,PhysRevB.58.9227}       &        &      209        &    0\textsuperscript{§}           &  &       175     &       &c&c&--&--  \\
\hline
\textcolor{blue}{${\rm UCu_2Si_2}$ (Tetra.)}  \cite{doi:10.1143/JPSJ.76.074708}       &        &      254        &    0\textsuperscript{§}           &  &       106     &   100    &c&c&1.75&20  \\
\hline
\textcolor{blue}{${\rm USb_2}$ (Tetra.)}  \cite{doi:10.1126/sciadv.aaw9061}       &        &      80        &    0\textsuperscript{§}           &  &       200     &       &c&c&1.9&27  \\
\hline\hline
\end{tabular}
\caption{A table of the coherence temperature $T^*_{\rm coh}$ defined as the maximum in resistivity; $T_{\rm cross}$, the temperature of the susceptibility crossing; $T_{\rm max}^{\rm hard}$ the temperature of the maximum in the low-temperature hard-axis susceptibility; $T_C$(K) or $T_N$(K) the Curie or Néel temperature; $x$ the doping fraction; P, pressure (GPa); The orientation of the magnetic moment (orien.); the high-temperature easy axis (HTEA); the size of the ordered moment, $\mu\ (\mu_B)$; and the Sommerfeld coefficient, $\gamma$ (${\rm mJmol^{-1}K^{-2}}$).\\
NM -- Non magnetic; NR -- Not resolved; prin. -- Principally\\ 
The superscript on $T^*_{\rm coh}$ values indicate the axis for which the resistivity measurement was taken.\\
$\ast$ the crossing occurs less than 1\,K above the Curie temperature.\\
$\dagger$ $T_{\rm cross}$ determined from mCW fits.\\
$\ddagger$ No coherence maximum observed.\\
§ No clear susceptibility crossing observed.\\
\P The coherence temperatures were determined from linear interpolations of resistivity measurements at different pressure values. 
}
\end{table*}

\onecolumngrid
\textcolor{blue}{\section{Derivation of the effective Hamiltonian in Eq. 7 }}

\color{blue}The variational parameters $\{u_{\mathbf{k}\alpha},v_{\mathbf{k}\alpha}\}$ are chosen to minimise $\langle \Psi |H_N | \Psi \rangle$ subject to the constraints on $\langle \hat{n}^f_{ai} \rangle$ and the total number of conduction electrons, which corresponds to solving
\begin{align}
    \frac{\delta }{\delta u_{\mathbf{k} \alpha}}\! \left( \langle H_N \rangle
    \!-\!\bar{\lambda}_1 \left( \left\langle \sum_i \hat{n}^f_{1i} \right\rangle \! -\!\frac{NN_s}{2} \right)\!-\!\mu  \left( \left\langle \sum_i \hat{n}^c_{i} \right\rangle - \frac{N N_s n_c}{2}\right)\right)
   \!=0,
\end{align}
where the Lagrange multipliers $\bar{\lambda}_1$ and $\mu$ are adjusted to give the expectation values $\langle \hat{n}^{f}_{1i}\rangle =\frac{N}{2}$ and $\langle \hat{n}_i^c\rangle = \sum_{ \alpha\sigma} \langle c^{\dagger}_{\alpha\sigma}(\mathbf{r}_i)c_{\alpha\sigma}(\mathbf{r}_i)\rangle =\frac{1}{2}n_cN$ .
The expectation value to be minimised can be computed using Wick's theorem and is equal to $\langle \Psi |H_{\rm MF}| \Psi \rangle$, where, up to a constant, $H_{\rm MF}=H_0+H_1$ is an appropriately chosen mean-field Hamiltonian with
\begin{eqnarray}
H_0&=& \sum_{\mathbf{k} \alpha\sigma} \Psi_{\alpha\sigma\mathbf{k}}^{\dagger}\begin{pmatrix}
        \epsilon_{\mathbf{k}} - \mu  & -J_K\bar{V}_1\\ -J_K\bar{V}_1&\bar{\lambda}_1
    \end{pmatrix} \Psi_{\alpha\sigma\mathbf{k}} + N N_s\left(J_K \bar{V}_1^2 +\frac{1}{2}\mu  n_c 
+\frac{1}{2}\bar{\lambda}_1  
\right) 
 -\sum_{i\alpha \eta} h^{\eta}_i\bar{S}^{\eta}_{2\alpha i}
,\nonumber\\
    H_1 &=& \sum_{i\alpha\eta} \left[(J_M\bar{s}^{\eta}_{\alpha i}+2D^{\eta}\bar{S}^{\eta}_{2\alpha i} -h^{\eta}_i) \hat{S}^{\eta}_{1\alpha}(\mathbf{r}_i) 
+(J_M\bar{S}_{1\alpha i}^{\eta}+J_M\bar{S}_{2\alpha i}^{\eta} - h^{\eta}_i)\hat{s}^{\eta}_{\alpha}(\mathbf{r}_i)
-J_M \bar{s}^{\eta}_{\alpha i} \bar{S}_{1 \alpha i}^{\eta}
    \right] 
    \nonumber\\
    &
\approx &
\sum_{i\alpha\eta} \left[(2D^{\eta}\bar{S}^{\eta}_{2\alpha i} -h^{\eta}_i) \hat{S}^{\eta}_{1\alpha}(\mathbf{r}_i) 
+(J_M\bar{S}_{2\alpha i }^{\eta} - h^{\eta}_i)\hat{s}_{\alpha}^{\eta}(\mathbf{r}_i)\right]
,
\end{eqnarray}
where $\Psi_{\alpha\sigma\mathbf{k}}^{\dagger}= \left(c_{\alpha\sigma}^{\dagger}(\mathbf{k}), f_{1 \alpha\sigma}^{\dagger}(\mathbf{k})\right)$, $\bar{V}_1=\langle \hat{V}_{1i} \rangle$, $\bar{S}_{a\alpha i}^{\eta}=\langle \hat{S}_{a\alpha}^{\eta}(\mathbf{r}_i) \rangle$, $\bar{s}^{\eta}_{\alpha i}=\frac{1}{2}\sum_{\sigma\sigma'}\langle c^{\dagger}_{\alpha\sigma}(\mathbf{r}_i) \sigma^{\eta}_{\sigma\sigma'} c_{\alpha\sigma'}(\mathbf{r}_i) \rangle$ and $J_M=2J_K/N$.  
The $|\Psi\rangle$ which minimises  $\langle \Psi |H_N| \Psi \rangle$ is the ground state of $H_{\rm MF}$. Working to quadratic order in $J_M,D^{\eta},h^{\eta}$,  we can treat $H_1$ perturbatively and use linear response theory to compute the ground state energy of $H_{\rm MF}$. Since $\bar{s}^{\eta}_{\alpha i}$ and $\bar{S}^{\eta}_{1\alpha i}$ are both of the order of $J_M,D^{\eta},h^{\eta}$, we have further approximated $H_1$.The change in the order parameters $\bar{V}_1,\bar{\lambda}_1$ and $\mu$ due to the perturbation is of second order in $J_M,D^{\eta},h^{\eta}$, and because they mark the stationary point of $\langle \Psi |H_0 | \Psi\rangle$, can be neglected.  We can now straightforwardly find the effective Hamiltonian
\begin{eqnarray}
H_{\rm eff} &=& -\frac{1}{2} \int_0^{\beta} \left\langle e^{H_0\tau}
H_1 e^{-H_0\tau}H_1
\right\rangle_{H_0} \; {\rm d}\tau
 -\sum_{i\alpha \eta} h^{\eta}\hat{S}^{\eta}_{2\alpha i}
 \nonumber\\
 &=&
 -\frac{1}{2}\sum_{ij \eta \alpha}
 \left[
 \chi_{ff}(\mathbf{r}_i-\mathbf{r}_j)\left(2D^{\eta}\right)^2\hat{S}^{\eta}_{2\alpha i} \hat{S}^{\eta}_{2\alpha j} 
 + 2\chi_{fc}(\mathbf{r}_i-\mathbf{r}_j)2D^{\eta}
 J_M \hat{S}^{\eta}_{2\alpha i}\hat{S}_{2\alpha j}^{\eta} 
 +\chi_{cc}(\mathbf{r}_i-\mathbf{r}_j)J_M^2\hat{S}_{2\alpha i}^{\eta}\hat{S}_{2\alpha j}^{\eta} 
 \right]
 \nonumber\\
 &&
 -\sum_{ij\alpha\eta}\left[\delta_{ij}\!-\! J_M(\chi_{cc}(\mathbf{r}_i-\mathbf{r}_j)+\chi_{cf}(\mathbf{r}_i-\mathbf{r}_j))\! -\!2D^{\eta}
    \left(
    \chi_{ff}(\mathbf{r}_i-\mathbf{r}_j) + \chi_{cf}(\mathbf{r}_i-\mathbf{r}_j)
    \right)\right] h^{\eta}_j \hat{S}_{2\alpha i},
\end{eqnarray}
where $\bar{S}^{\eta}_{2\alpha i}$ have been promoted to spin-$1/2$ operators, constant terms have been dropped, $\chi_{ff}(\mathbf{r})=\int_0^{\beta} \langle e^{H_0\tau}  \hat{S}^{\eta}_{1\alpha}(\mathbf{r}) e^{-H_0\tau} \hat{S}_{1\alpha}^{\eta}(\mathbf{0})\rangle_{H_0}$, $\chi_{cf}(\mathbf{r})=\int_0^{\beta} \langle e^{H_0\tau}  \hat{s}^{\eta}_{\alpha}(\mathbf{r}) e^{-H_0\tau} \hat{S}^{\eta}_{1\alpha}(\mathbf{0})\rangle_{H_0}$, $\chi_{cc}(\mathbf{r})=\int_0^{\beta} \langle e^{H_0\tau}  \hat{s}^{\eta}_{\alpha}(\mathbf{r}) e^{-H_0\tau} \hat{s}^{\eta}_{\alpha}(\mathbf{0})\rangle_{H_0}$, and $\chi(\mathbf{q})=\sum_{i}\chi(\mathbf{r})e^{-i\mathbf{q} \cdot \mathbf{r}_i}$. These susceptibilities can be easily evaluated to obtain the expressions given in the main text once $H_0$ is expressed in terms of the heavy-fermion quasiparticle creation/annihilation operators: $\sum_{\nu \mathbf{k}\alpha\sigma} E^{\nu}_{\mathbf{k}} \left( u_{\mathbf{k}}^{\nu} c_{\sigma\alpha}(\mathbf{k})+v_{\mathbf{k}}^{\nu} f_{1\sigma\alpha}(\mathbf{k})\right)^{\dagger}\left( u_{\mathbf{k}}^{\nu} c_{\sigma\alpha}(\mathbf{k})+v_{\mathbf{k}}^{\nu} f_{1\sigma\alpha}(\mathbf{k})\right)=\sum_{\nu \mathbf{k}\alpha\sigma} E^{\nu}_{\mathbf{k}} d^{\dagger}_{\nu \alpha\sigma}(\mathbf{k}) d_{\nu \alpha\sigma}(\mathbf{k})$, where

\begin{eqnarray}
    u_{\mathbf{k}}^{\pm} = 
    \frac{E^{\pm}_{\mathbf{k}\alpha} +\bar{\lambda}_{1}}{\sqrt{(E^{\pm}_{\mathbf{k}} +\bar{\lambda}_{1})^2 + J_K^2 \bar{V}_1^2}},\quad
       v_{\mathbf{k}}^{\pm} = 
    \frac{-J_K\bar{V}_1}{\sqrt{(E^{\pm}_{\mathbf{k}} +\bar{\lambda}_{1})^2 + J_K^2 \bar{V}_1^2}},
    \quad
    E^{\pm}_{\mathbf{k}}\!\! = \frac{
    \epsilon_{\mathbf{k} } -\mu - \bar{\lambda}_{1} 
     \pm \sqrt{ \left(\epsilon_{\mathbf{k} } -\mu + \bar{\lambda}_{1} 
    \right)^2 + 4J_K^2 \bar{V}_1^2}}{2}.
\end{eqnarray} 

\textcolor{blue}{\section{Metamagnetism}}
We mention here the apparent correlation between metamagnetism and the crossing of uniform susceptibilities. As shown in Table I, compounds that exhibit metamagnetism along the high-temperature easy axis (HTEA) also exhibit a crossing of the uniform susceptibilities at a temperature $T_{\rm cross}$, as well as a maximum along the HTEA at a temperature $T^{\rm hard}_{\rm max}<T_{\rm cross}$.

\end{document}